\documentclass[twocolumn,prb,nofootinbib]{revtex4}
\usepackage{graphicx}
\usepackage{epsfig}
\usepackage{dcolumn}
\usepackage{bm}
\usepackage{amsmath}

\begin{document}

\title{Entanglement and errors in the control of spins by optical coupling}

\author{G. F. Quinteiro} \author{C. Piermarocchi}

\affiliation{Department of Physics and Astronomy, Michigan State
University, East Lansing, Michigan 48824}

\date{\today}

\begin{abstract}
We analyze the optical quantum control of impurity spins in proximity
to a quantum dot. A laser pulse creates an exciton in the dot and
controls the spins by indirect coupling. We show how to determine the
control parameters using as an illustration the production of maximal
spin entanglement.  We consider errors in the quantum control due to
the exciton radiative recombination. The control errors in the
adiabatic and nonadiabatic case are compared to the threshold needed
for scalable quantum computing.
\end{abstract}

\maketitle


\section{Introduction}
Impurity spins embedded in semiconductors are currently under
investigation for quantum computing implementations. Recently, optical
techniques have been proposed to control the spin-spin coupling and
realize two-qubit quantum
gates.~\cite{piermarocchi02,piermarocchi04,stoneham03,rodriquez04,ramon04,nazir04,pazy03}
The optical method suggests the possibility of an ultrafast control of
the qubits. The flexibility in the control that can be obtained by
pulse shaping~\cite{chen01} and the absence of noisy contacts
represent additional advantages.  On the experimental side, ensemble
optical measurements have demonstrated the production of spin
entanglement for impurities embedded in a semiconductor
host.~\cite{bao03} More recently, the measurement of the quantum state
of a single impurity spin obtained by coupling it to a single exciton
in a quantum dot (QD) has been experimentally carried
out.~\cite{besombes04} In this paper we study theoretically the
control of impurity spin states when the interaction among them is
controlled by optically-generated excitons in a QD. We analyze the
control errors due to the radiative recombination of the exciton that
mediates the interaction between the spins. Moreover, we illustrate
how the control parameters can be obtained directly from simple
analytical expressions. The method is applied to design the control
parameters in the production of maximal spin entanglement.

\section{system}
\begin{figure}
\includegraphics[scale=0.25]{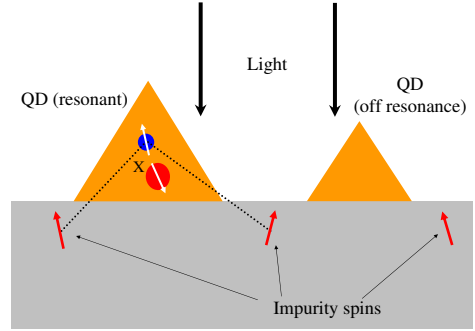}
\caption{Scheme the system: localized spins located near self
assembled QDs are coupled by an exciton created by a laser pulse. Dots
of different size provide selective control and
readout.\label{fig_Sys}}
\end{figure}

The physical system consists of two impurity spins placed close or
inside a QD in such a way that there is not a direct interaction
between them.  A schematic picture is given in Fig. \ref{fig_Sys}. By
coding the qubit in more than one spin efficient schemes for
fault-tolerant~\cite{bacon00} and exchange-only~\cite{divincenzo00}
quantum computation can be naturally applied to this setup. Dots of
different size provide the frequency selectivity to address specific
spin pairs and realize two-qubit readouts.  The model we use contains
few parameters describing the exciton-light and exciton-impurity
coupling and can be applied to different physical systems. For
instance, it can be used for excitons localized by monolayer
fluctuation in III-V and II-VI quantum wells and interacting with a
finite number of localized impurities as in
Ref.~\onlinecite{bao03}. III-V or Si/Ge self-assembled QDs can also be
used as shown in Fig.~\ref{fig_Sys}. For typical semiconductor systems
we can restrict the analysis to heavy-hole excitons due to the
splitting between heavy-hole and light-hole bands in the dot. The
heavy-hole exciton spans a four dimensional space consisting of two
optically-active and two dark states. We treat the interaction between
the electromagnetic field and the excitons semiclassically, and we
consider spin states that interact only with the photoexcited electron
in the dot.  This is the case for instance of donor impurity spins in
typical semiconductors because the electron-hole exchange is much
smaller than the electron-electron exchange. Notice that by using
circularly polarized light the exciton induces, besides the spin-spin
coupling, also a local effective magnetic field on the
spins.\cite{piermarocchi04,combescot04} This effective magnetic field
can be controlled by the laser polarization and disappears for
linearly polarized light. We will consider below the case of
circularly polarized light.

The exciton-spin part of the Hamiltonian can be written as
\begin{equation}
H_0=\epsilon N - 2V (J^2-S^2-L^2)
\end{equation}
where $S$ is the total spin of the two impurities ($\hbar=1$) and
$\epsilon$ is the energy of the exciton in the dot.  The operators $N$
and $L$ are defined as $N = b^\dag_{\uparrow}b_{\uparrow} +
b^\dag_{\downarrow}b_{\downarrow}$, $L_x =\frac {1} {2}
(b^\dag_{\downarrow} b_{\uparrow} + b^\dag_{\uparrow}b_{\downarrow})$,
$L_y = \frac {i} {2} (b^\dag_{\downarrow}b_{\uparrow} -
b^\dag_{\uparrow}b_{\downarrow})$ and $L_z = \frac {1} {2}
(b^\dag_{\downarrow}b_{\downarrow} - b^\dag_{\uparrow}b_{\uparrow})~.$
$b^\dagger_\downarrow$ creates an optically active exciton with
electron spin $-1/2$ and hole spin $+3/2$, while $b^\dagger_\uparrow$
creates a dark state exciton with electron spin $+1/2$ and hole spin
$+3/2$. The total angular momentum $J_i=L_i+S_i$, and $V$ is the
exchange interaction between the impurity spins and the photoexcited
electron in the dot.  The strength and the sign of $V$ depend on the
system. For instance, this coupling is expected to be ferromagnetic
for electrons in the dot interacting with localized rare-earth
magnetic impuritites, while it is antiferromagnetic for a dot
mediating the interaction between shallow
donors.~\cite{piermarocchi04,ramon04} Without loss of generality, we
will assume $V>0$ below. The coupling of the excitons in the QD and
the external laser field is given by
\begin{equation}
\label{HLX} H^{(LX)}=\frac{\Omega(t)}{2} e^{-i\omega t}
b^\dag_\downarrow + hc
\end{equation}
where $\Omega(t)$ is the time-dependent Rabi energy associated with
the optical pulses, and $\omega$ is the energy of the laser. We
consider only anti-clockwise polarization ($\sigma_+$) which generates
excitons with electron and hole spin states $-1/2$ and $+3/2$,
respectively. Excitons with hole spin $-3/2$ are not included in the
model since they are not excited by $\sigma_+$ light and the impurity
spins can only flip the spin of the photoexcited electron.

\begin{figure}
\includegraphics[scale=0.25]{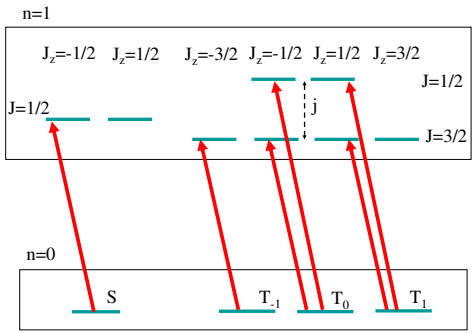}%
\caption{Energy level diagram and optical selection rules for
$\sigma_+$ polarized light. In $n=1$, the total splitting is $j= 6 V$.
\label{levels}}
\end{figure}

A scheme of the relevant energy levels is given in
Fig. \ref{levels}. In the ground state $n=0$ we have the singlet and
the triplet states corresponding to the two non-interacting
impurities. In the excited state $n=1$, the electron in the dot splits
the triplet states in a quadruplet $J=3/2$ and a doublet $J=1/2$. The
total Hilbert space is thus spanned by a total of $12$ states. The
arrows in the scheme identify the selection rules for $\sigma_+$
optical transitions. The transitions have different oscillator
strengths, which are calculated using the Clebsch-Gordan coefficients.
Notice that the light does not connect directly states with different
spin $S$.  The structure of the energy levels provides a natural
readout scheme for the coded logical qubit $|0_L\rangle$, $|1_L
\rangle$ in the exchange-only scheme.~\cite{divincenzo00} An optical
setup similar to the one for single spin readout~\cite{besombes04}
could be used: a single peak at $\epsilon$ corresponds to
$|0_L\rangle$ while two peaks separated by $6 V$ correspond to the
logical state $|1_L\rangle$.

\section{quantum control}
In order to illustrate how to design the optical control we consider
the production of maximal spin entanglement. We choose the the initial
state $|\uparrow\downarrow\rangle\otimes|0\rangle$ as the tensor
product of a linear superposition of impurity states $\{|S
\rangle,|T_0 \rangle \}$, and the exciton $|0\rangle$ representing an
empty QD. We consider separately the case of infinite and finite
$\gamma^{-1}$, i.e. spontaneous radiative recombination lifetime for
the exciton in the dot. In the first case we determine analitically
the control parameters that provide maximal spin entanglement. In the
second case, we solve numerically the master equation for the full
system in Fig.~\ref{levels}.  This will allow us to analyze errors due
both to the radiative recombination and to the finite probability of
remaining with an exciton in the dot at the end of a pulse. The latter
is an error similar to a double occupancy error in the case of spins
controlled by gate voltages \cite{hu00}.  Ideally, the QD must be
empty at the end of each optical pulse, and this can be achieved by an
adiabatic evolution, or by a nonadiabatic evolution plus additional
conditions in the pulse area~\cite{kaplan04}.

\subsection{Infinite radiative lifetime} 
\begin{figure}
\centerline{\includegraphics[scale=.8]{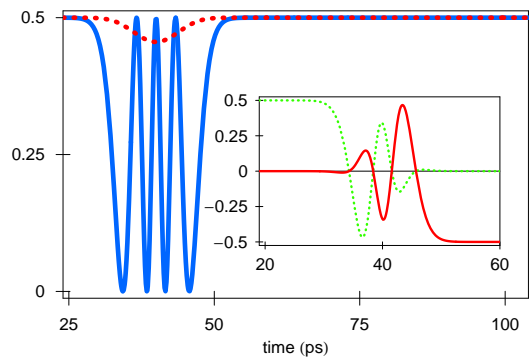}}
\caption{(Color online) Nonadiabatic control, $\gamma=0$.
Evolution of the $|S\rangle$ (solid blue line) and $|T_0\rangle$
(dashed red line) populations under a Gaussian pulse of area
$8\pi$. The temporal width of the pulse $\tau$ is $7.02$ ps and
the ratio $\Omega/V$ is 0.6697. (Inset) Real (dashed green line)
and imaginary (solid red line) part of the coherence $\langle
S|\rho|T_0\rangle$.\label{reso}}
\end{figure}

We first
analyze the ideal case of a nonadiabatic evolution at $\gamma=0$.
We call nonadiabatic the evolution that follows from a laser
resonant with at least one transition between the $n=0$ and $n=1$
subspaces in Fig.~\ref{levels}. This implies that there is a
substantial exchange of energy between the electromagnetic field
and the dot, which in turn results in a significant population
inversion during the pulse. Using a numerical simulation we
illustrate in Fig.~\ref{reso} the evolution of the $|S\rangle$ and
$|T_0\rangle$ populations under a Gaussian pulse giving a Rabi
energy of the form,
\begin{equation}
\Omega(t)=\frac{\Omega}{\sqrt{\pi}}e^{-(t/\tau)^2}~.
\label{rabit}
\end{equation}
The pulse is resonant with the bare exciton energy, which in the
scheme of Fig.~\ref{levels} corresponds to a resonant transition for
the singlet state. In order to have no excitonic population at the end
of the pulse, we need the pulse area for the resonant excitation to be
multiple of $2\pi$, therefore $\Omega$ and $\tau$ are chosen so that
the pulse area is $\Omega \tau=8\pi$. Notice that the population of
the ground state singlet $|S\rangle$ is completely depleted during the
pulse but at the end comes back to the original population ($0.5$). In
contrast, the triplet ($|T_0\rangle$) population follows an adiabatic
evolution due to the exchange interaction affecting the optical
resonance. In Fig.~\ref{reso}~(inset) we show the real and imaginary
part of the coherence $\langle S|\rho|T_0\rangle$. In order to
create the maximally entangled state we need a $\pm\pi/2$ phase in
this matrix element and the chosen optical pulse achieves this
goal. This relative phase transforms, for example, the state
$|\uparrow \downarrow \rangle$ into $2 ^{-1/2}(|S\rangle+
i|T_0\rangle) \propto |\uparrow \downarrow \rangle + i|\downarrow
\uparrow \rangle$. For a given value of the exchange coupling $V$ and
pulse width $\tau$, the maximum intensity of the field $\Omega$ in
Eq.~(\ref{rabit}) is found from the roots of the equation
\begin{equation}
\phi_T(\Omega,V,\tau)\pm \pi/2=0
\label{phase}
\end{equation}
where $\phi_T=\int_{-\infty}^{\infty}\lambda^0_T(t)~dt$ is the dynamic
phase that the state $|T_0\rangle$ picks up following the adiabatic
evolution. Notice that since the pulse is a multiple of $2\pi$ the
singlet will only pick up a trivial phase ($\pm 1$).  $\lambda^0_T$ is
the eigenvalue satisfying $\lambda^0_T(\pm\infty)=0$ for a 3-level
Hamiltonian representing the triplet states,
\begin{equation}
H_T(t)=\frac{1}{2}\left[
\begin{array}{ccc}
0&\sqrt{\frac{2}{3}}\Omega(t)&\sqrt{\frac{1}{3}}\Omega(t)\\
\sqrt{\frac{2}{3}}\Omega(t)&2\delta-\frac{2}{3} j&0\\
\sqrt{\frac{1}{3}}\Omega(t)&0&2\delta+ \frac{4}{3} j
\end{array} \right]~.
\label{ht}
\end{equation}
The optical detuning $\delta=\epsilon-\omega$ and $j=6V$ is the
splitting in the excited state between $J=3/2$ and $J=1/2$ states. If
we assume that the three eigenvalues of the matrix in Eq.~\ref{ht} do
not cross during the pulse evolution, the expression for
$\lambda^0_T(t)$ can be written as
\begin{equation}
\lambda^0_T(t)=\frac{j}{9}+\frac{2}{3}\delta+
\frac{q(t)}{3}\cos\left(\frac{\theta(t)}{3}\right)
\label{lam}
\end{equation}
where
\begin{equation}
\theta(t)=2k\pi+\arccos\left(\frac{r(t)}{q^3(t)}\right)
\label{theta}
\end{equation}
with
\begin{eqnarray*}
q(t)&=&\sqrt{28 j^2+12j\delta +36\delta^2+27 \Omega^2(t)} \\
r(t)&=&4 (4 j- 3\delta)(5j + 3\delta)(j+ 6\delta)- 81
(2j+3\delta)\Omega^2(t)~.
\end{eqnarray*}
 If the exciton impurity coupling is
ferromagnetic ($j>0$), we have to take in Eq.~(\ref{theta}) $k=1$ for
$\delta > j/3$, $k=2$ for $-2j/3 < \delta < j/3$, and $k=3$ for
$\delta < -2j/3$. In contrast, for $j<0$, we have to take in
Eq.~(\ref{theta}) $k=1$ for $\delta > -2j/3$, $k=2$ for $1/3 j< \delta
< -2j/3$, and $k=3$ for $\delta < j/3$. The analytical expression in
Eq.~(\ref{lam}) allows us to determine exactly the control parameters
from the roots of Eq.~(\ref{phase})~.
\begin{figure}
\centerline{\includegraphics[scale=.8]{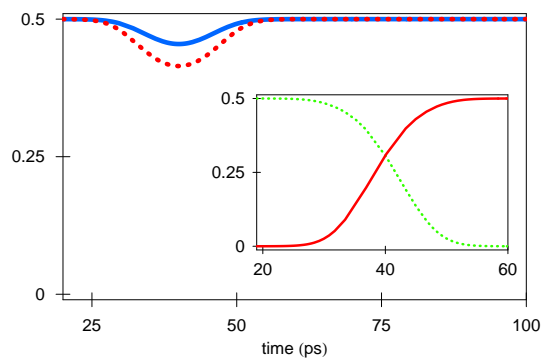}}
\caption{(Color online) Adiabatic control, $\gamma=0$.  The laser
is tuned in $2$ meV below the bare excitonic energy. Evolution of
the $|S\rangle$ (solid blue line) and $|T_0\rangle$ (dashed red
line) populations under a Gaussian pulse. The temporal width of
the pulse $\tau$ is $10.2$ ps and the ratio $\Omega/V$ is 1.24.
(Inset) Real (dashed green line) and imaginary (solid red line)
part of the coherence $\langle S|\rho|T_0\rangle$.\label{offres}}
\end{figure}

In the $\gamma=0$ adiabatic regime the laser pulse is tuned away
from the optical resonances between the $n=0$ and $n=1$ levels. An
example of a simulation of an adiabatic control is shown in
Fig.~\ref{offres}. The laser is tuned $2$ meV below the bare
excitonic energy and $1$ meV below the triplet resonances
corresponding to $J=3/2$ in Fig.~\ref{levels}.  We plot in
Fig.~\ref{offres} the same quantities of Fig.~\ref{reso}. Notice
that in this case the pulse area can be arbitrary, provided the
adiabaticity is preserved. The change of phase in the coherence
$\langle S|\rho|T_0\rangle$ is now obtained with a smooth
transition. The control parameters in this adiabatic case are
determined by the roots of
\begin{equation}
\phi_T(\Omega,V,\tau)-\phi_S(\Omega,V,\tau)\pm \pi/2=0~.
\label{sineq}
\end{equation}
In contrast to the case of Fig.~\ref{reso}, the singlet now picks up a
nontrivial dynamic phase
$\phi_S=\int_{-\infty}^{\infty}\lambda^0_S(t)~dt~$ where $\lambda^0_S$
is the eigenvalue of the singlet Hamiltonian
\begin{equation}
H_S(t)=\frac{1}{2}\left[
\begin{array}{cc}
0&\Omega(t)\\ \Omega(t)&2 \delta
\end{array} \right]
\label{hs}
\end{equation}
with the property $\lambda^0_S(\pm \infty)=0$. As for $H_T$ this has a
simple analytical form $\lambda^0_S(t)=\frac{\delta}{2} \pm
\frac{1}{2}\sqrt{\delta^2+\Omega^2(t)}~$, ($+$ for $\delta<0$ and $-$
for $\delta>0$ ) which can be used to determine the control parameters
from the roots of Eq.~(\ref{sineq}).

\subsection{Finite radiative lifetime} In order to
determine how this control scheme is affected by the finite
lifetime of the exciton in the dot we introduce a finite value for
$\gamma$, and solve the master equation $\dot{\rho} = -i[H, \rho]
+ \cal{L}[\rho]$ using the values of the control parameters
corresponding to the evolution of Figs.~\ref{reso} and
\ref{offres}. $\cal{L}[\rho]$ is the Liouvillian superoperator
that can be written as ${\cal{L}}[\rho]=L^\dagger \rho L
+\frac{1}{2}\{L^\dagger L, \rho \}$ where $L^\dagger
=\sqrt{\gamma} ~b_\downarrow$ accounts for the spontaneous
radiative recombination of the exciton in the dot. Once $\rho$ is
obtained, a $4 \times 4$ reduced density matrix for the impurity
spins $\rho_S$ is computed by tracing out the exciton degrees of
freedom. The entanglement in the Bell state is mostly sensitive to
decoherence processes and its analysis provides a good test for
the scheme. We quantify the error on the reduced density matrix
$\rho_S$ using two different methods, the Purity and the Peres
criterion of separability.~\cite{peres96} According to the Peres
criterion a state is entangled iff $E_{min}<0$, where $E_{min}$ is
the minimum eigenvalue of a matrix constructed by transposing the
non-diagonal $2\times2$ blocks of $\rho_S$. A maximally entangled
state has a $E_{min}=-1/2$. The deviation from that value gives a
measure of the effect of the radiative recombination on the
entanglement and we quantify the entanglement error as $\Delta E=
E_{min}+1/2$. The purity of $\rho_S$ is a different parameter that
characterizes the error in the spins states due to their
entanglement with the exciton in the dot. We quantify this error
as $\Delta P=Tr{\rho_S^2}-1$. In principle there are errors that
can disentangle the spin states without a change in the purity,
for instance by affecting the phase picked up in
Eq.~(\ref{sineq}). Therefore, in principle the errors induced by
$\gamma$ affect independently $\Delta E$ and $\Delta P$.

\begin{figure}
\centerline{\includegraphics[scale=1]{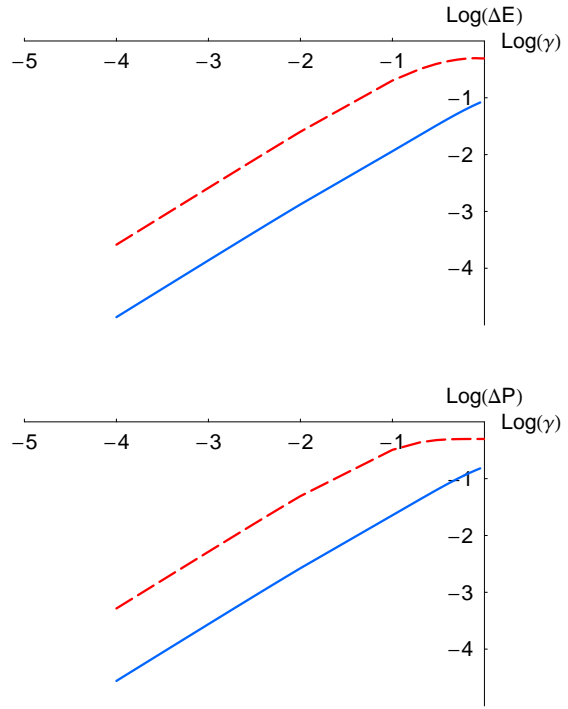}} \caption{(Color
online) Log-Log plot of the deviation from maximal entanglement
(upper panel) and maximal purity (lower panel) as a function of
the radiative recombination $\gamma$ (in meV). Solid blue line:
adiabatic evolution. Dashed red line: nonadiabatic evolution.
\label{gamma}}
\end{figure}
We compare in Fig.~\ref{gamma} the errors in the Entanglement
$\Delta E$ (upper) and in the Purity $\Delta P$ as a function of
$\gamma$ for both the adiabatic and nonadiabatic evolution. Both
errors increase linearly at small $\gamma$. However, the errors in
the adiabatic case are always smaller than in the nonadiabatic
case in the range of parameters we have investigated. We remark
that, due to the incommensurability of the eigenvalues of $H_S$
and $H_T$, there are not special conditions that would give
perfect entanglement with square pulses as in the case of a direct
spin-spin coupling.~\cite{kaplan04} An important figure of merit
for the application of this quantum control technique to quantum
computation is provided by the {\it error per gate} parameter.
This has to be below a threshold value in order to make scalable
quantum computing possible.  The estimate for such a threshold
depends on assumptions on the error model and device capabilities
but the $10^{-4}$ value ~\cite{gottesman97} is usually used as a
benchmark in typical experimental implementations.  The error in
the entanglement production gives an estimation of the error per
gate since the quantum operation done corresponds to a
$\sqrt{SWAP}$ modulo some single qubit operations. We see in
Fig.~\ref{gamma} that the $10^{-4}$ threshold can be achieved for
$\gamma$ smaller than 1$\mu eV$. Self assembled QDs have typically
a ground state exciton lifetime of the order of one or more
nanoseconds and would reasonably be in this region of parameters.

\section{Conclusions}
In conclusion, we have analyzed the entanglement production
between two spin-impurities induced by an exciton in a neighboring
quantum dot. In the case of $\gamma=0$, the parameters for the
quantum control can be analytically determined from the roots of
simple integral equations. We showed that the finite lifetime
$\gamma^{-1}$ of the exciton in the dot can affect the purity of
the spin states and introduces errors in the entanglement
production. In addition we found that such errors increase
linearly with $\gamma$ and can be kept below the $10^{-4}$
threshold for error correction if parameters typical of self
assembled QDs are used in the simulation.

\acknowledgments This work was supported by the NSF Grant
DMR-0312491. We thank Prof. T. A. Kaplan for discussions.

\end{document}